\documentclass[11pt]{scrartcl}
\usepackage[pdftex]{graphicx}
\title{
Detector Utility for International Linear Collider
}
\author{
Yasuhiro Sugimoto\footnote{Talk presented at the International Workshop
on Future Linear Colliders (LCWS2019),  Sendai, Japan, 28 October -- 1 November, 2019. C19-10-28.}
\vspace{2cm}\\
High Energy Accelerator Research Organization (KEK) \\
Tsukuba, Ibaraki 305-0801, Japan
\vspace{1cm}
\date{}
}
\begin{document}
\maketitle
\begin{abstract}
Detectors for experiments at International Linear Collider require utilities
such as electricity and cooling water, as well as the space to locate the 
relevant service facilities. 
In this article, a possible design of caverns for utilities and 
services near the interaction point is presented. 
The design has been made intending
to satisfy requirements from detectors, as well as requirements
from the accelerator.
\end{abstract} 

%
\section{Introduction}
In order to operate detectors for experiments at International Linear Collider (ILC),
supply of utilities such as electricity and cooling water is indispensable. 
Because ILC detectors are located about 100~m underground,
transformers for electricity and heat exchangers and water pumps 
for cooling water have to be located underground.
There are other apparatus for detector services which should be
put near detectors. 
However, there should be some distance between
some of these apparatus and detectors because vibration caused
by these apparatus should not affect the beam alignment at the interaction point (IP).
From these considerations, utility/service caverns are necessary
at an appropriate distance from detectors.

Concerning the location of the utility/service cavern, there have been several designs 
so far~\cite{adolphsen, miyahara, hayano, sugimoto}.
The design in TDR \cite{adolphsen} is obsolete now because the baseline design
of the detector hall (DH) and the tunnels around DH 
has been largely modified from TDR. The design given in \cite{miyahara} is
the present baseline design, but the utility/service cavern in this design  is not satisfactory
because its location is asymmetric with respect to the position of the two detectors.
In the design by Tohoku team~\cite{hayano}, the length of DH is extended by 25~m,
and this extended part is used as a utility/service cavern.
This utility/service cavern is occupied by utilities for the accelerator only,
and there is no space for detector utilities. 
A counter-proposal was made
by the detector side~\cite{sugimoto} in which separated utility caverns for
the detectors and the accelerator are proposed. 
However, in this counter-proposal,
high voltage (66~kV) power lines for the whole accelerator system have to cross
the damping ring, which is not favorable for operation of the accelerator.

In this article, a compromised design of the utility/service cavern (USC) 
which is acceptable for both the accelerator side and the detector side is proposed. 
The design is based on surveys of requirements from sub-detector groups of 
ILD~\cite{behnke} which is one of the two detectors for ILC experiments.

\section{Outline of the compromised design}
\begin{figure}
\centering
\includegraphics[scale=0.52]{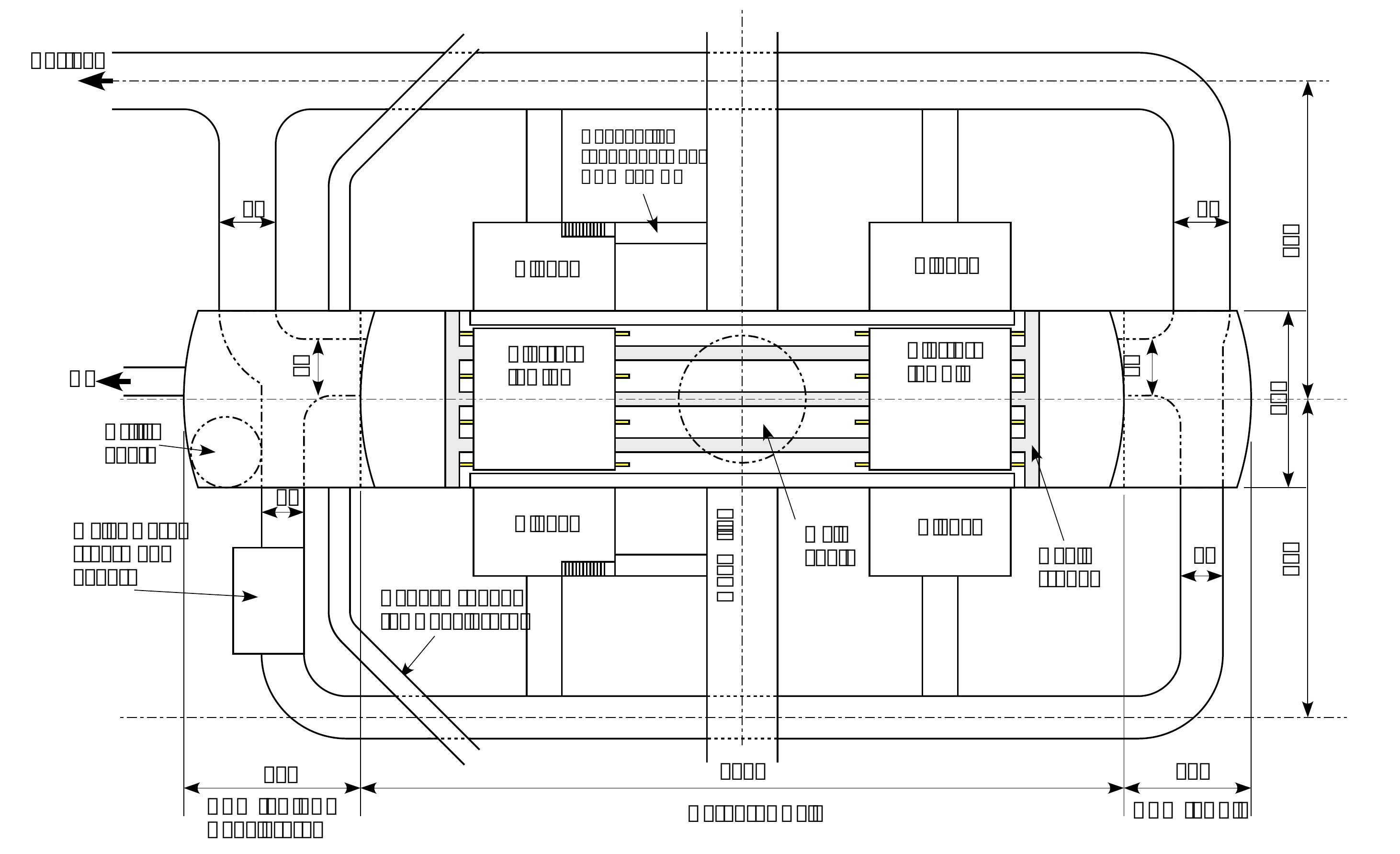}
\hspace*{0.3cm}
\includegraphics[scale=0.52]{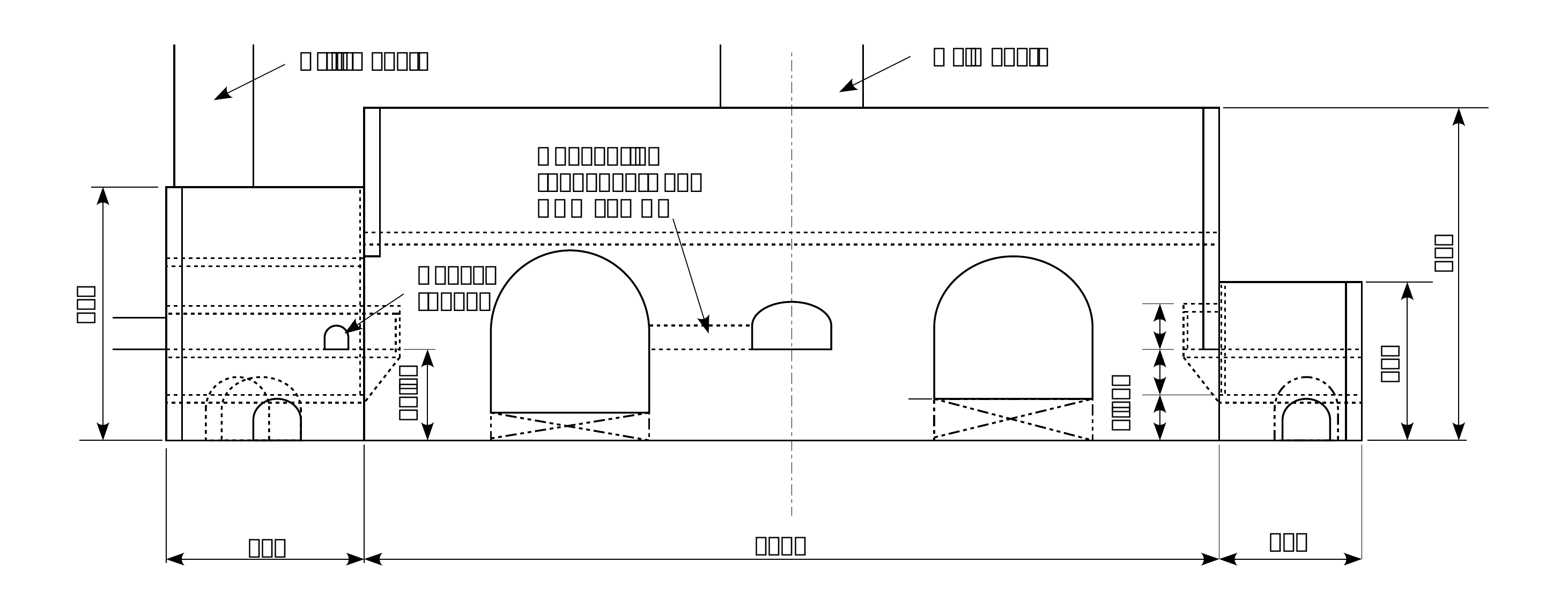}
\caption{Plan view (top) and side view (bottom) of the compromised design 
of DH and USC.}
\label{caverns}
\end{figure}

Plan view and  side view of the caverns of the 
compromised design are shown in
Figure~\ref{caverns}.
In this design,
USC is attached to  DH, and a vertical shaft (utility shaft) is
connected to USC. Accelerator bypass tunnels  connect
USC to the accelerator tunnels. 
The electricity of 66~kV  and the cooling water for the accelerator are distributed
from the surface through the utility shaft, USC, and the bypass tunnels.
USC is also connected to the damping ring tunnel by a short tunnel
to provide  electricity and cooling water for the damping ring.
The design described above is similar to that of the Tohoku design.
However, configuration in USC is modified from the Tohoku design
to secure enough space for ILD utilities. An additional USC for 
another detector SiD is attached to DH on the opposite side to
keep symmetry with respect to the two detectors.
The electricity of 6.6~kV and the cooling water are supplied
from the utility shaft to USC for SiD through 6~m wide tunnel next to DH.

\begin{figure}
\centering
\includegraphics[scale=0.6]{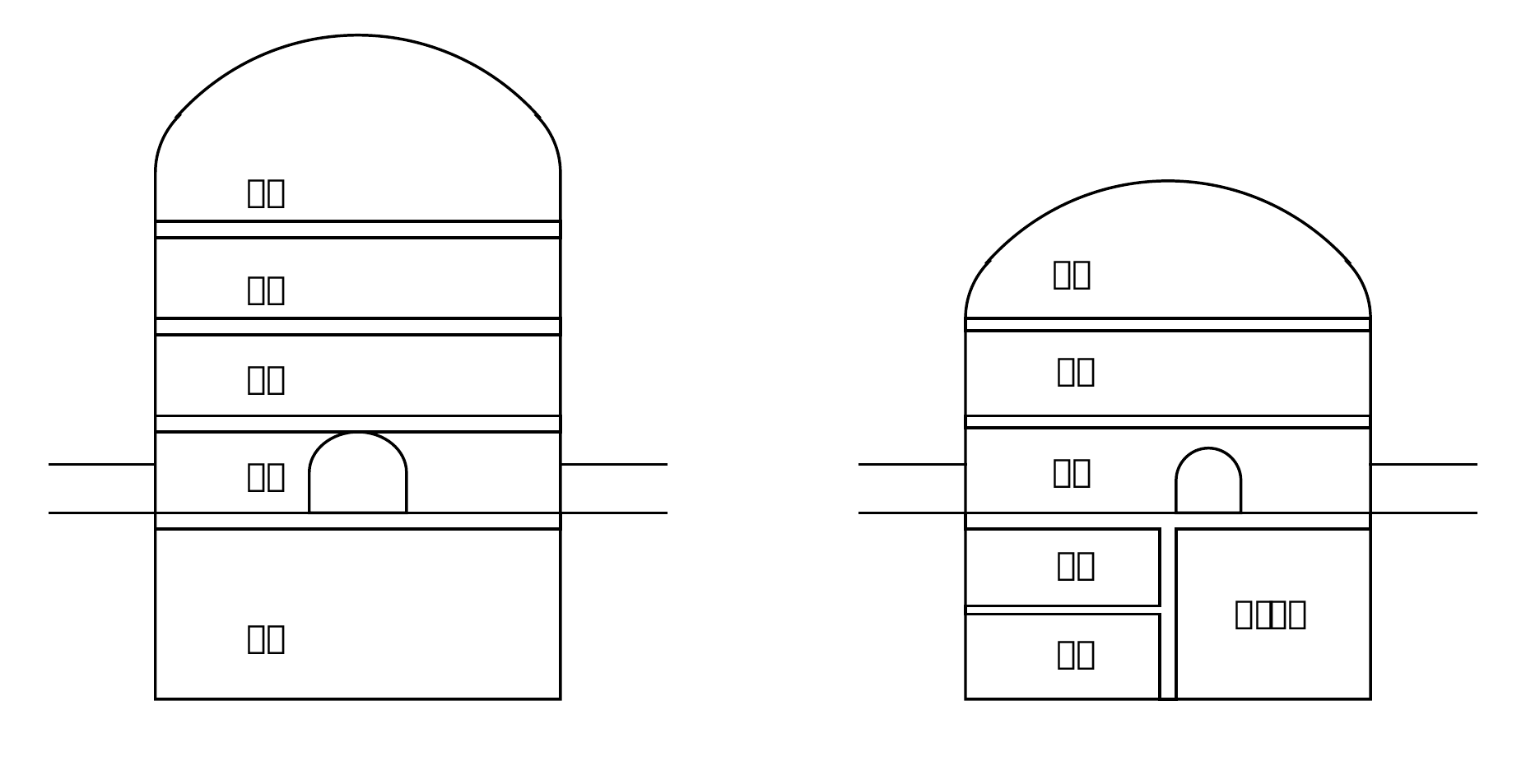}
\caption{Structure of USC in the Tohoku design (left) and 
in the compromised design (right) seen from the DH side.}
\label{USC}
\end{figure}

In the compromised design, configuration in USC is modified from 
the Tohoku design as shown in Figure~\ref{USC}.
The height of the USC is reduced from the Tohoku design
in order to reduce the total excavation volume.
Usage of each floor is modified as shown in Table~\ref{floor}.
In the compromised design, 
air handling units (AHUs) of the heating, ventilation, and air-conditioning
(HVAC) system are located in a  surface building.
The waste water treatment system should be put in another
small cavern. By such re-arrangement, enough space for
utilities and services for ILD is secured.

 \begin{table}
 \centering
 \caption{Configuration in USC for accelerator and ILD.}
 \begin{tabular}{ | l | l | l | l |} 
 \hline
 \multicolumn{2}{| l |}{Tohoku design} &  \multicolumn{2}{| l |}{Compromised design} \\
 \hline
 5F & HVAC & & \\
 \hline
 4F & HVAC & 5F & Gas systems for ILD \\
   & & & Laser system for ILD \\
   & & & Workshop \\ 
 \hline  
 3F & Cooling water system for accelerator & 4F & Cooling water system for accelerator \\
 \hline
 2F & Transformers for accelerator & 3F & Transformers for accelerator \\
     &  & & Transformers for ILD \\
 \hline
 1F & Helium cryogenics & 2F & Cooling systems for ILD \\ \cline{3-4}
     & Water treatment system &  & Helium cryogenics for QF1/CC \\
     &  & 1F & Cooling systems for ILD \\
     &  &    & Cooling water system for ILD \\    
 \hline
 \end{tabular}
 \label{floor}
 \end{table}

USC is separated from DH by a wall up to its ceiling except for the entrance
of DH so that helium gas blowing out from detector solenoids in case of
a severe accident should not harm the people working on upper floors of USC.

Service galleries will be built on the wall of DH at the same levels of floors
of USC. These galleries work as  egresses from USC, as well as the maintenance space
of cables and pipes between detectors and USC. 
In addition, there are service galleries along the wall between USC and DH on 3F and 4F.
These two floors of the service gallery should be load bearing because
electronics racks and power supplies for magnets will be put on these floors. 
 
\section{Requirements for utilities and services from detector side}
In order to operate detectors for ILC experiments, the following 
utilities and services are required:
\begin{itemize}
  \item Electricity; 400~V ($3\phi$), 200~V ($3\phi$, $1\phi$), 100~V ($1\phi$);
  \item Cooling water; Chilled water, normal temperature water, low conductivity water; 
  \item Heating, ventilation, and air-conditioning (HVAC) system;
  \item Compressed helium gas for cryogenics;
  \item Sub-detector cooling systems;
  \item Gas systems for sub-detectors;
  \item Laser system for alignment;
  \item Electronics (19-inch) racks for data acquisition;
  \item Compressed air for air-pads. 
\end{itemize}  

High voltage electric power is lowered to 6.6~kV on the surface, and sent to
underground USC through the utility shaft. In USC the 6.6~kV electricity  is 
stepped down to 400/200/100~V by transformers (cubicles),
and fed to power supplies for detectors.
Requirements for electric power have been surveyed in ILD group~\cite{sugimoto2019Feb}.
About 1~MW is necessary for underground facilities in case of ILD.
In addition, about 3~MW would be necessary for surface facilities.
These numbers have to be doubled if another detector SiD is
taken into account. The space for the cubicles for ILD would be
about 7~m$\times$4~m including surrounding space for
maintenance and safety.

Most of the dissipated power is taken away by cooling water. 
The cooling water is supplied from the surface through the utility shaft. 
High pressure due to height difference of  $\sim 100$~m
between the surface and USC is 
cut by heat exchangers.
There are three loops of the cooling water in the underground caverns;
chilled water, normal temperature water, and low conductivity water (LCW).
The chilled water is used for fan-coil units and sub-detector cooling systems.
The normal temperature water is used for sub-detector cooling systems.
The LCW is used to cool the magnet power supply.
For redundancy, we will use two sets of a heat exchanger and a pump for each loop.

Similarly to most of the LHC experiments, AHUs for HVAC are 
located in a surface building, and the air is supplied and extracted 
through air-ducts in the main shaft just above the interaction point.

Compressed helium gas for the detector solenoids and the final quadrupole
magnets (QD0) is supplied from the surface to DH through the main shaft.
The cold box for the solenoid magnet and QD0 will be put on the
detector itself or on the detector platform.
Compressed helium gas for QF1 and the crab cavity (CC) will be 
supplied through either the utility shaft or the main shaft. 
The cold box for QF1/CC
is placed on 1F of USC~\cite{okamura}.

Each sub-detector has its own cooling system. 
Sub-detector cooling systems could occupy quite a large area~\cite{verlaat}.
Therefore, we assume 5~m$\times$3~m area for each sub-detector cooling system.
There will be 11 sub-detectors for ILD.
Some of the sub-detector cooling systems require their location
at the level of 1F.

Some of the sub-detectors of ILD use gasses for their operation. 
The gas control systems are located in USC, while the 
gas bottles are store in a hut on the surface.
The laser system is necessary for tracker alignment.
A hut for the laser system which can be locked for the safety reason
should be in USC.
 
The  electronics racks for data acquisition  can be placed on 
 3F of the service gallery  and on 5F of USC.
 We need a space for more than 50 racks in these places.
 
The compressed air is necessary for air-pads and some tooling.
The air-pads are apparatus which can lift  heavy load using pressurized air,
 and we can move the heavy load with very small friction.
 The ILC detectors share one interaction point alternatively
 in a push-pull operation scheme. For that purpose, each detector
 is put on a platform which is a thick slab made of reinforced concrete.
 In order to move the platform together with the detector,
 many air-pads are used.
 ILD uses another set of air-pads to move the detector 
 (open and close the end-cap of the detector) on the platform.
 For the operation of these air-pads, a high pressure air compressor 
 or buffer tanks
 should be placed in USC.
 
 In the compromised design shown in Figure~\ref{caverns} and Figure~\ref{USC}, 
 we can find enough space to satisfy the requirements
 described above.
 Concerning USC for SiD, the total floor area available for detector utilities
 is larger than $\mathrm{500~m^2}$. 
 
 \section{Excavation volume}
 Compared to the baseline design given in~\cite{miyahara}, 
 the total excavation volume of caverns and tunnels of the Tohoku design 
 is about $4200~\mathrm{m^3}$ larger because of the large volume of USC,
 and that of the compromised design
 is about $5600~\mathrm{m^3}$ larger
 because  new caverns for SiD utilities and for waste water treatment system are added. 
 However, this difference can be compensated by reducing 
 the height of DH by $2~\mathrm{m}$.
 In the DH designs in \cite{adolphsen, miyahara}, 5~m space of straight wall
 is reserved above the crane rails. This space is necessary for the crane of 400~t 
 capacity which was necessary at the time of TDR~\cite{adolphsen}.
 In the recent baseline design of DH, the crane capacity in DH is 40~t.
 In this case, 3~m space above the crane rails is required. 
 Therefore, the DH height can be reduced by 2~m.
 This option of reducing the height of DH should be considered seriously.
  
 \section{Transportation between the surface and underground caverns}
The  ILC detectors are supposed to be integrated in the assembly hall
 on the surface just above  the main shaft. Each detector (ILD or SiD) consists of
 several big pieces, 
 and is lowered into DH through the main shaft piece by piece using a 4000~t gantry crane.
 Most of the sub-detectors are installed to these pieces in the assembly hall before lowering.
 However, some of the sub-detectors such as the main tracker and the
 inner trackers could be installed in DH after lowering depending on the 
 assembly schedule.
 After completion of the detector assembly, some sub-detectors could require repairment
 and  have to be carried out form  DH.
 In these cases, these sub-detectors may have to be
transported between DH and the surface 
through the horizontal access tunnel. 
The USC 1F serves as a
part of the access tunnel. The largest sub-detector which
could have to be transported through the access tunnel is the
 central tracker TPC in case of ILD. The diameter of TPC itself is 3.6~m. 
We need a frame with suspension mechanism  to absorb the vibration
during the transportation. Therefore, we assumed 4.6~m width of the load,
and studied if such a load can be transported into or out from DH
in the design shown in Figure~\ref{caverns}. Because there is
unreachable area of the crane in DH, the trailer has to go well inside of  DH.
Using a simulation program
of locus of trailer trucks, we confirmed that there is no problem
in the transportation of TPC.

Apparatus in USC are transported using an elevator in  the utility shaft.
The heaviest and largest equipment to be transported is a
$300~\mathrm{kVA}$ transformer unit (cubicle).
A typical weight is 1.3~t, and the height is 2.2~m.
Therefore, the elevator with the capacity of 2~t and the
height of the door of 2.5~m would work.
The weight of the power supply for the detector solenoid is not clear yet.
However, it can be placed on the 4F service gallery from the DH side
using a mobile crane.

\section{Utility requirements during construction period}
Because the campus of ILC near the interaction point (IP campus) is built
in countryside, a new high voltage (154~kV) electricity power line
of about 15~km long has to be built to the IP campus~\cite{hayano}. 
Water for ILC will be
supplied from the north end of the main linac (ML), and fed to 
cooling water systems in other places
through the ML tunnel~\cite{hayano}. The piping in the ML tunnel  to IP
will be finished at the end of 7th year (Y7) or later from 
the ground breaking~\cite{adolphsen}.
On the other hand, we assume the detector construction at the IP campus 
starts in late Y3~\cite{sugimoto2016}.
Therefore, utilities at the IP campus will be fully available after Y8, 
and limited during detector construction period. We have to clarify how much utilities
is necessary during the construction period. If necessary, we should think about
complementary power source such as a solar power system, or
utilization of another remote campus in a area where the utilities are already available.

\section{Summary} 
A new design of USC near the interaction point of ILC combined with DH
has been presented in this article. In this design, USC is attached to DH on both ends, 
and has the size of 25~m wide, 25~m long, and 32~m high for ILD/accelerator,
and  25~m wide, 18~m long, and 20~m high for SiD.
It has been confirmed that there is enough space in USC for ILD and accelerator.
The total floor area of USC for SiD is larger than $\mathrm{500~m^2}$.
Because we don't have enough inputs from SiD group, 
more discussion on the requirements from SiD is necessary
to make a more detailed design of USC for SiD.
 If the height of DH is reduced by 2~m, the total excavation volume 
of DH, USC, and tunnels is similar to that of the baseline design.
There is no problem in transportation of sub-detectors between the surface 
and DH.
The elevator in  the utility shaft should have capability of carrying cubicles into USC, 
i.e. 2~t capacity and 2.5~m door height.
Utility requirements at IP campus during detector construction period
should be clarified, and some complementary power source or utilization
of remote campus should be  considered if necessary.

\begin{footnotesize}

\end{footnotesize}


\end{document}